\documentclass[12pt]{iopart}

\usepackage{graphicx}

\newcommand{\be}{\begin{eqnarray}}
\newcommand{\ee}{\end{eqnarray}}

\begin{document}

\title[]{The second act of hydro: small perturbations}

\author{Pilar Staig and Edward Shuryak$^*$}

\address{Department of Physics and Astronomy, State University of New York,
Stony Brook, NY 11794}
\ead{Edward.Shuryak@stonybrook.edu}
\begin{abstract}
Hydrodynamical description of the ``Little Bang" in heavy ion
collisions is surprisingly successful: here  we
systematically study  propagation of small perturbations 
treated hydrodynamically.  
Using analytic description of the expanding fireball known as the  ``Gubser flow", we proceed to linearized equations for perturbations.
As all variables are separated and all
equations  solved (semi)analytically, we can
 collect  all the harmonics and
reconstruct the complete Green function of the problem, even in the viscous case.  Applying it to the power spectrum we  found acoustic minimum at the $m=7$ and maximum at $m=9$, which remarkably 
have some evidence for both in the data. We estimate effective viscosity and size of the perturbation from a fit to power spectrum.
The shape of the two-point correlator is also reproduced remarcably well. At the end we argue that independent perturbations are local,
and thus harmonics phases are correlated.
\end{abstract}
\footnote{$^*$ The speaker}

The actual talk was based on  papers \cite{Shuryak:2009cy,Staig:2010pn,Shuryak:2011vf,Staig:2011wj,Staig:2011as} and included
discussion of the ``jet edge", correlations between harmonics and other general ideas, in particular the relation to Cosmology.
Due to lack of space, this written version focuses narrowly  on the specific results, due to astonisingly good agreement between our calculations and wide range of data
presented at this historic conference.

A correct theory usually provides  impressive higher order applications; recall e.g. Newton's explanation of tides or the higher order QED corrections.
The title indicate a new round of applications of hydrodynamics, now for small perturbations of the  exploding fireball.
Since this development is new, 
the first application should be as simple as possible. Thus, we only consider
(near)central nuclear collisions, in which the impact parameter is small enough to be neglected and thus there is no average elliptic flow.
The  zeroth order system (without fluctuations) is thus assumed to be completely axially symmetric.
We use analytic rather than numerical methods.  Therefore, 
on a technical level, we base it on the analytic solution for central collisions of conformal plasma objects, to be referred to as the ``Gubser flow", see \cite{Gubser}.  Obviously here we cannot give technical details about this solution: let us just define the comoving coordinates $\rho,\theta$ given by:
$\sinh{\rho} = -\frac{1-q^2\tau^2+q^2r^2}{2q\tau},
\tan{\theta} = 
\frac{2qr}{1+q^2\tau^2-q^2r^2}
$
a combination of the transverse distance $r$ and the proper time $\tau=\sqrt{t^2-z^2}$, with $t,z$, being the lab. time and the longitudinal (beam) coordinate.
The solution does not depend on the longitudinal rapidity and azimuthal angle $\phi$. The parameter $q$ represents the system's size. 
 The equation for the linearized first-order perturbations around this solution for the temperature perturbation $T = T_0(1+\delta) $ can be written in closed form, for zero viscosity it is 
\begin{eqnarray}
& &\frac{\partial^2 \delta}{\partial \rho^2} -
\frac{1}{3\cosh^2{\rho}} \left( \frac{\partial^2 \delta}{\partial
\theta^2}  +\frac{1}{\tan{\theta}}\frac{\partial \delta}{\partial
\theta}+ \frac{1}{\sin^2{\theta}}\frac{\partial^2 \delta}{\partial
\phi^2}
\right)+\frac{4}{3}\tanh{\rho}\frac{\partial \delta}{\partial \rho}=0
\label{T_pert_eqn}
\end{eqnarray}
for longer version with viscosity see \cite{Staig:2011wj}. 
 These equations allow for separation of variables: in fact the bracketed part is nothing else but the angular part of the spherical Laplacian, and thus it is solved by the usual spherical harmonics $Y_{lm}(\theta,\phi)$ and harmonics depend on $l$ rather than $m$. The $\rho$ part of the equation is
treated as the ``time dependence".

 Each harmonic can be viewed as an independent oscillator, transferring the energy between its potential (pressure) and kinetic (flow) forms, with certain $l$-dependent frequency. We studied the $\rho$ (time) evolution of the
perturbations, with and without viscosity and found that the amplitude is greatly reduced for higher harmonics in the viscous case, 
while the oscillation phase is more or less preserved. 
Both in the Big and Little Bangs, the time allocated to the hydrodynamical stage of the evolution is limited by the  ``freezeout time" $\tau_f$, after which the collision rates  can no longer keep up with the expansion. At this time  each harmonic has   a different  phase of its oscillation. In the Universe the temperature fluctuation is just read from the sky, and thus nodes of  $\delta_l(fo)$ correspond to these minima. In the Little Bang one has to calculate the specific combination of the temperature and flow perturbations. This includes the calculation of how the freezeout surface is modified: but  the nodes/maxima of this ``observable" combination  generate the minima/maxima in the power spectrum.  
   This simple physics is very robust, the minima/maxima are easily predictable and rather insensitive to many details such as dissipation. 

\begin{figure}[t!]
\begin{center}
\includegraphics[width=10. cm]{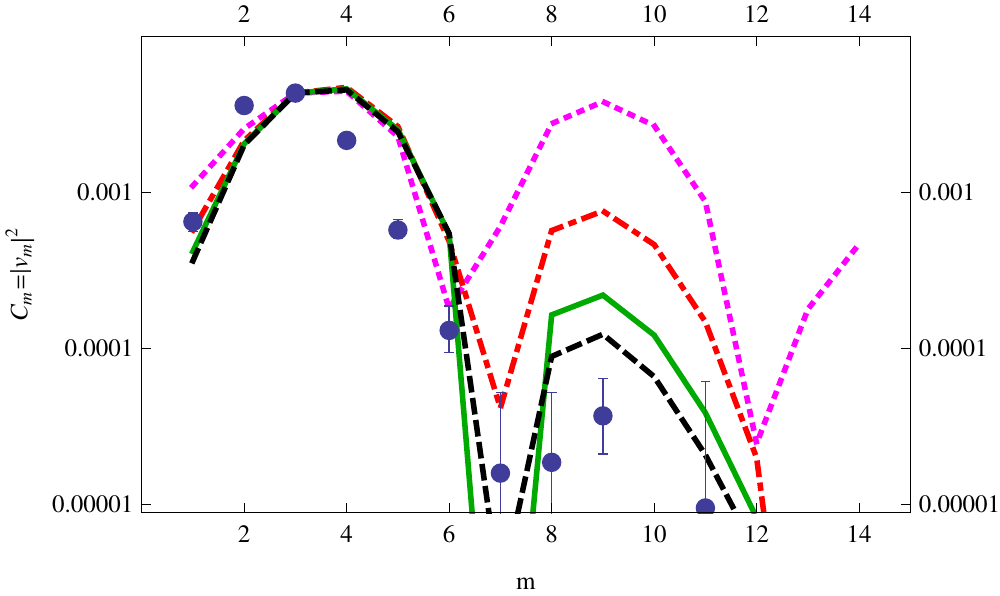}
\end{center}
\vspace{-5ex}\caption{(Color online) 
The power spectrum of flow harmonics, $|v_m|^2$ versus m. The data points are preliminary ATLAS data  \cite{ATLAS_corr}. Four curves top-to-bottom (dashed magenta,dash-dotted red, solid green and dashed blue)
are our calculations for viscosity-to-entropy ratios $4\pi\eta/s=0,1,1.68,2$, respectively. All curves are normalized to the m=3 harmonics.
 }\label{fig_spectral}
\end{figure}

   Fig.\ref{fig_spectral} shows how this idea works in practice. 
   We assumed that the size of the initial pertubation
is 0.5 fm, corresponding to the Glauber collision model, and use different viscosities as shown below.
     Our calculation predicts the acoustic minimum at $m=7$  and the  maximum at $m=9$. Quite remarkably the very first data
on this power spectrum, from ATLAS collaboration, show the same minimum and maximum.
   One can further see, that the data points for the largest nonzero harmonics $m=6,9$ are incompatible with zero viscosity on the level of many sigmas, and even the ADS/CFT value is clearly excluded. The  
estimated viscosity from this plot is about twice larger  than the AdS/CFT prediction, namely
$  {4\pi \eta \over s}\approx 2 $.   
   
\begin{figure}[!t]
\begin{center}
\includegraphics[width=5 cm]{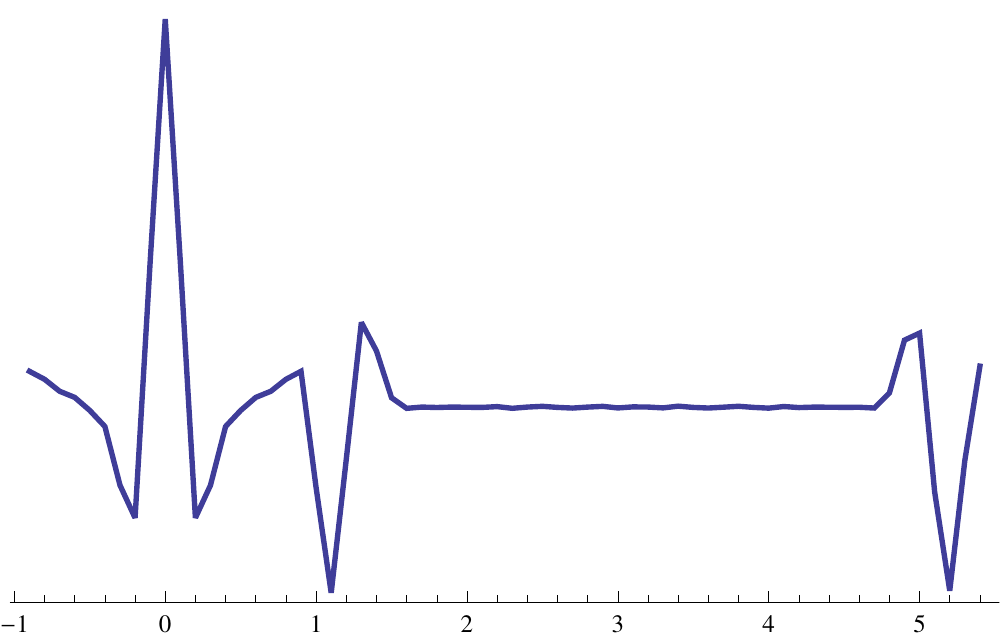}
\includegraphics[width=5 cm]{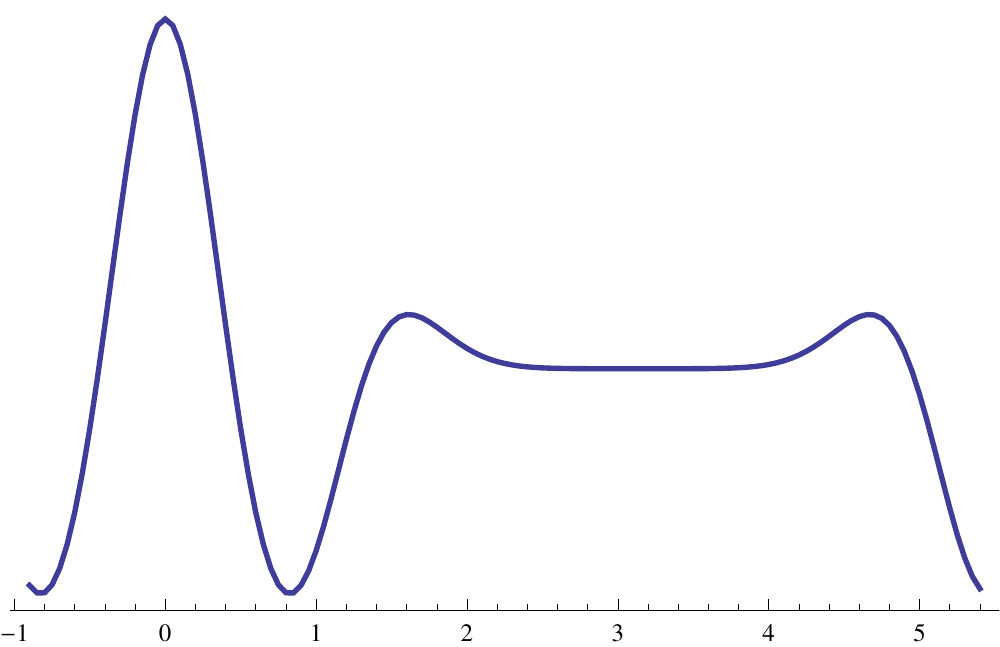}
\includegraphics[width=5. cm]{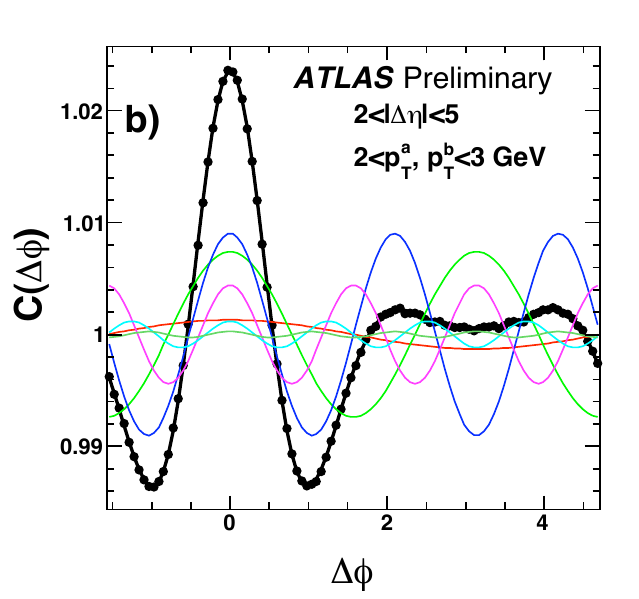}
\end{center}
\vspace{-5ex}\caption{The two-pion distribution as
a function of azimuthal angle difference $\Delta\phi$, our calculation \cite{Staig:2011wj}. Two left plots are
for viscosity-to-entropy ratios $\eta/s=0,\,0.134$, respectively, while the right one is from  ATLAS report \cite{ATLAS_corr}. Similar data 
 have been reported at this meeting by all RHIC/LHC collaborations.
}\label{2pdist}
\end{figure}

 A traditional plot is the two-point  correlation function as a function of the relative angle. Fig.\ref{2pdist} shows its calculated shape, with and without viscosity,
to be compared with the experimental data. Once again,  with an appropriate viscosity,  the agreement of the shape becomes amazingly good.
   Note in particular the width of the main peak and the hight of a plateau. Moreover,
    nothing has been fitted to the data: it is the first calculation with rather approximate speed of sound and the freezeout time preselected in the calculation
    supeimposed with the first round of data from this conference.
(The delta-function perturbation has been placed at some typical location $r=4.1 \, fm$: how it depends on that is discussed in \cite{Staig:2011wj}
in details.) 

  Two opposite scenarios, from incoherent to fully coherent, can be defined as follows:\\
  (i) The  ``minimal Gaussian model" is a possibility that the perturbations are in fact incoherent noise, fully characterized by the mean squares  $<\delta_l(in)^2>$. Apparent agreement with the Green function calculation appears simply because of weak dependence of   $<\delta_l(in)^2>$ on $l$.\\
  (ii) The  ``maximal coherence models", assumes that harmonics are coherent and the physical perturbations are in fact  described by our Green function. The physical argument for it is that quantum fluctuations in nucleon-nucleon interactions at different locations (in the transverse plane) should to a large extent be independent from each other.  Phase correlations has  been demonstrated
  in  \cite{Staig:2010pn} based on the Glauber model.  

    How can one  tell the difference between them experimentally? By going beyond the two-point correlators.  Indeed, as argued in our paper \cite{Staig:2010pn}, the three-point correlators allow to measure the ``resonance terms" including three harmonics such that $m_1+m_2+m_3=0$, and extract the terms with the combinations of phases such as
$cos(m_1\psi_1+m_2\psi_2+m_3\psi_3)$. For central collisions, for which the two-point correlators are $\sim 10^{-3}$, one would expect the three-point ones to be another factor 30 or so down in magnitude. For non-central collisions one may use large value of the ellipticity, provided one of the harmonics involved is the second one (e.g. $m_1=2$) and less statistics be needed: but in this case the theory is more complex and yet needs to be developed. 

   We end up, reminding the reader, that while the calculations themselves are quite technical,
   the underlying physics is very simple, basically the sound circles similar to what one finds while throwing stones into the pond.
 The sound velocity 
$c_s\sim 1/2$ and the time till freezeout $\tau_{FO}\sim 2R$ (where R
is the nuclear size), thus the maximal radius of the
circles (the ``sound horizon") is simply $H_s\sim R$. Thus, in terms of the azimuthal angle $\Delta \phi=\pm 1 radian$.
Strong radial flow dramatically enhances the contrast at the edge of the fireball, making small
deviations of the freezeout surface into two ``horns"  \cite{Shuryak:2009cy}, best seen
 in a specially tuned region of the secondaries, with $p_t\sim 2-3 \,GeV$ where the radial flow effects are at its maximum.
    The circle and horns are well seen in our calculation.  They were found
    in hydrodynamical simulations by the Brazilian group \cite{Andrade:2009em}, who went on and provided a simple explanation for the three-peak structure of the correlation functions. Indeed, if one calls the two horns + and -, the 4 combinations
for the particle pairs are + +,- -,+ - and - +. The first two produce a peak near zero relative angle and the last two would correspond to the relative angle being $twice$ the (angle projected) sound horizon, or about 2 radians. This is indeed what is experimentally observed.
We thus expect that the relative phases of the harmonics be such
as to {\em cancel}  the (non-existing) third horn of the leading $m=3$ harmonics. It is still amazing to find, that the pictures holds against the data up to the 9-th
harmonics!
  
   Partly an inspiration for this work is clearly  the events in Cosmology
during the last decade, which gave us  observations of ``the frozen sound" scale, also known as the sound horizon, both in the   cosmic microwave background (CMB) radiation and in the distribution
of galaxies. 
As these observations  turn Cosmology into a much more quantitative science, we hope  their analogues for the ``Little Bang" will also fix the global parameters in question much better.

\end{document}